\documentclass[final,5p,times,twocolumn]{elsarticle}

\usepackage{amssymb}
\usepackage{amsmath}

\biboptions{sort&compress}

\usepackage[colorlinks=true,allcolors=blue]{hyperref}

\journal{Physics Letters A}

\begin{document}

\begin{frontmatter}

\title{Universal semiclassical drift velocity of magnetic edge channels}

\author[first,third]{Ekaterina R. Sleptsova}
\author[second,third]{Alexey A. Sokolik \corref{cor1}}
\ead{asokolik@hse.ru}
\cortext[cor1]{Corresponding author}
\affiliation[first]{
            organization={National Research Nuclear University MEPhI},
            addressline={Kashirskoe Shosse 31},
            city={Moscow},
            postcode={115409}, 
            country={Russia}}
\affiliation[second]{
            organization={Institute for Spectroscopy, Russian Academy of Sciences},
            addressline={Fizicheskaya 5}, 
            city={Troitsk, Moscow},
            postcode={108840}, 
            country={Russia}}
\affiliation[third]{
            organization={National Research University Higher School of Economics},
            addressline={Pokrovsky Bulvar 11}, 
            city={Moscow},
            postcode={109028}, 
            country={Russia}}

\begin{abstract}
Magnetic edge channels, arising at the edge of a two-dimensional electron gas in a magnetic field, are topologically protected and propagate unidirectionally along the edge. Their group velocities at the Fermi level, being averaged harmonically, result in drift velocity, which is related to the edge density of states, quantum capacitance, and the velocity of edge magnetoplasmons. Using semiclassical quantization of electron skipping orbits, we show that the drift velocity equals 1/2 of the electron Fermi velocity in the limit of a large number of edge channels crossing the Fermi level. This result, being universal for a hard edge and for isotropic dispersion, is checked numerically for massive electrons and for massless electrons in graphene. The drift velocity at smooth electrostatically defined edges lacks such universality. In the case of anisotropic dispersion, the semiclassical drift velocity at the hard edge reduces to a coordinate-velocity correlation along the cyclotron orbit.
\end{abstract}


\end{frontmatter}

\section{Introduction}
\label{introduction}

Landau levels of a two-dimensional electron gas (2DEG) in a perpendicular magnetic field exhibit an upward bending when electron wave functions approach the edge due to the confining action of the latter \cite{Halperin1982}. The resulting magnetic edge channels circulating around the sample in a definite direction determined by the magnetic field orientation were recognized later as topologically protected edge states, whose number at the Fermi level equals the number of completely occupied bulk Landau levels \cite{Hatsugai1993}.

Group velocities of magnetic edge channels at the Fermi level play a key role in quantized conductance \cite{Buttiker1988}, quantum capacitance \cite{Christen1996,Bosco2017}, and other phenomena, including heat transport \cite{Sivan1986,Hu2009,Bocquillon2013} and photocurrent generation \cite{Plank2019,Singh2024,Mikhailov2026}. Several papers \cite{Wassermeier1990,Volkov1991,Talyanskii1992,Balev1997,Mikhailov2001,Hashisaka2013,Bosco2017,Lin2021,Wang2023,Sokolik2024} drew the connection between this velocity (or related quantum capacitance of the edge) and the drift velocity of edge magnetoplasmons \cite{Mast1985,Volkov1988,Mikhailov2001}, which adds to their ``electromagnetic'' velocity caused by electron Coulomb repulsion and Hall currents in a magnetic field. 
It was emphasized in Ref.~\cite{Sokolik2024}, that whenever several magnetic edge channels cross the Fermi level, the harmonic mean of their group velocities has the meaning of the edge magnetoplasmon drift velocity (although arithmetic averaging was also suggested \cite{Volkov1991,Mikhailov2001,Petkovic2014}). Such harmonic averaging of group velocities is essential for calculating the conductance of one-dimensional systems in the multichannel regime \cite{Buttiker1985}.

\begin{figure}[t]
	\centering 
	\includegraphics[width=0.9\columnwidth]{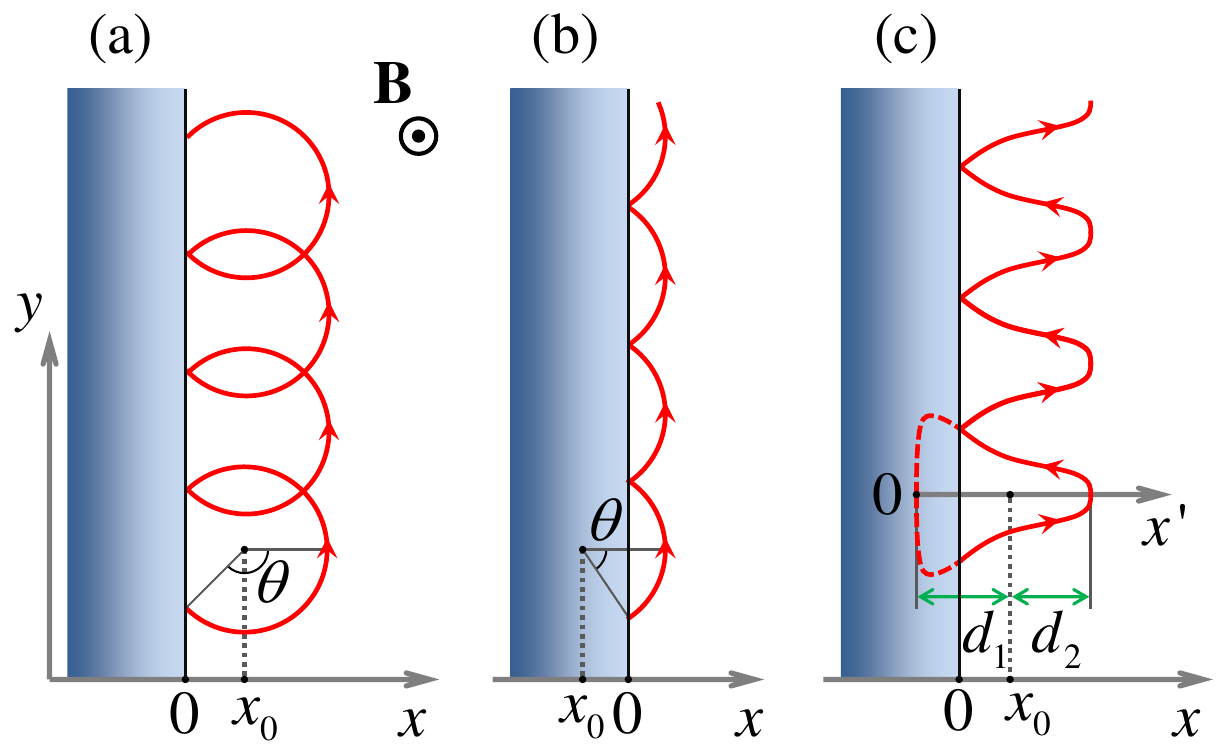}	
	\caption{(a,b) Skipping orbits in magnetic field for the coordinates of guiding centers (a) $x_0>0$ and (b) $x_0<0$. An electron is allowed to move only in the region $x>0$, and $\theta$ is the half span angle of the orbit. (c) Skipping orbit of arbitrary shape, mirror symmetric in the $y$ direction. The horizontal dimensions of the full cyclotron orbit are $d_1$ and $d_2$ to the left and to the right from its reference point $x_0$.}
	\label{Fig1}
\end{figure}

The drift velocity of edge magnetoplasmons was evaluated on the basis of experimental data on semiconductor quantum wells \cite{Delgard2021}, graphene \cite{Petkovic2013,Petkovic2014,Kumada2014}, and quantum anomalous Hall insulators \cite{Martinez2024}. It is generally believed that the drift velocity of magnetic edge channels $v_\mathrm{D}$ should be slightly smaller than the electron Fermi velocity $v_\mathrm{F}$, and the experiments confirm this expectation. In a classical picture, the magnetic edge channels are formed by the skipping orbits of electrons moving along circles in a magnetic field and bouncing off the edge (Fig.~\ref{Fig1}); thus their drift velocity (average velocity along the edge) should indeed be comparable to the Fermi velocity (linear velocity of the cyclotron rotation).

In this paper, we resolve the issue of the correct drift velocity value, which is defined as the harmonic mean of group velocities of magnetic edge channels crossing the Fermi level. We show that for a hard edge this velocity attains the universal value $v_\mathrm{D}=\frac12v_\mathrm{F}$ in the semiclassical limit when a large number of channels cross the Fermi level. This result is obtained by calculating and averaging the group velocities in the semiclassical approximation for arbitrary dispersion laws of electrons traversing the skipping orbits, and is applicable for any isotropic monotonous dispersion. Quantum calculations of the drift velocity for 2DEG of massive electrons and for massless electrons in graphene (at zigzag and armchair edge orientations) demonstrate that $v_\mathrm{D}$ oscillates as a function of the Fermi level around the universal value $\frac12v_\mathrm{F}$. The arithmetic mean of group velocities attains the value $v_\mathrm{D}\approx0.62v_\mathrm{F}$ in the semiclassical limit for isotropic electron dispersions. Meanwhile, the drift velocity at the smooth edge of 2DEG, \cite{Wassermeier1990,Talyanskii1992,Hashisaka2013}, which depends on the shape of the confinement potential, is generally not related to $v_\mathrm{F}$. We also present the calculation of the drift velocity for general anisotropic dispersion laws, when the cyclotron orbits have mirror symmetry relative to the edge normal, and obtain a closed-form expression for this velocity.

The paper is organized as follows. In Sec.~\ref{Sec2}, we present the quantum-mechanical analysis of the magnetic edge states for 2DEG and graphene, followed by the general calculation of the drift velocity and the analysis of its properties. Sec.~\ref{Sec3} describes semiclassical calculations of the drift velocity for isotropic and anisotropic dispersion laws resulting in the closed-form expressions. In Sec.~\ref{Sec4} we verify numerically the predicted universal results for the drift velocity of 2DEG and graphene with zigzag and armchair edges, and then state our conclusions and discuss possible future research directions.

\section{Quantum-mechanical analysis}
\label{Sec2}

\subsection{2DEG}

For electrons moving in the $(x,y)$ plane in a perpendicular magnetic field $\mathbf{B}=\{0,0,B\}$ it is convenient to choose the Landau gauge of a vector potential $\mathbf{A}=\{0,Bx,0\}$. An electron wave function $\psi$ satisfying the Schrodinger equation $(2m)^{-1}(-i\hbar\nabla+e\mathbf{A}/c)^2\psi=E\psi$ (electron charge is $-e$, where $e>0$) can be factorized as $\psi(x,y)=L_y^{-1/2}e^{iky}\phi(x)$, with $\phi(x)$ obeying the Schrodinger equation for a displaced harmonic oscillator:
\begin{equation}
-\frac{\hbar^2}{2m}\phi''(x)+\frac{m\omega_\mathrm{c}^2}2\left(x-x_0\right)^2\phi(x)=E\phi(x),\label{Schrodinger}
\end{equation}
where $\omega_\mathrm{c}=eB/mc$ is the cyclotron frequency, and
\begin{equation}
x_0=-l^2k\label{x0}
\end{equation}
is the $x$-coordinate of the oscillation center, $l=\sqrt{\hbar c/eB}$ is the magnetic length. Assuming the finite size of the sample $L_y$ in the $y$ direction and applying periodic boundary conditions $\psi(x,0)=\psi(0,L_y)$, one obtains the allowed values of the wave number $k$ in the $y$ direction quantized in the units of $2\pi/L_y$.

\begin{figure}[t]
	\centering 
	\includegraphics[width=\columnwidth]{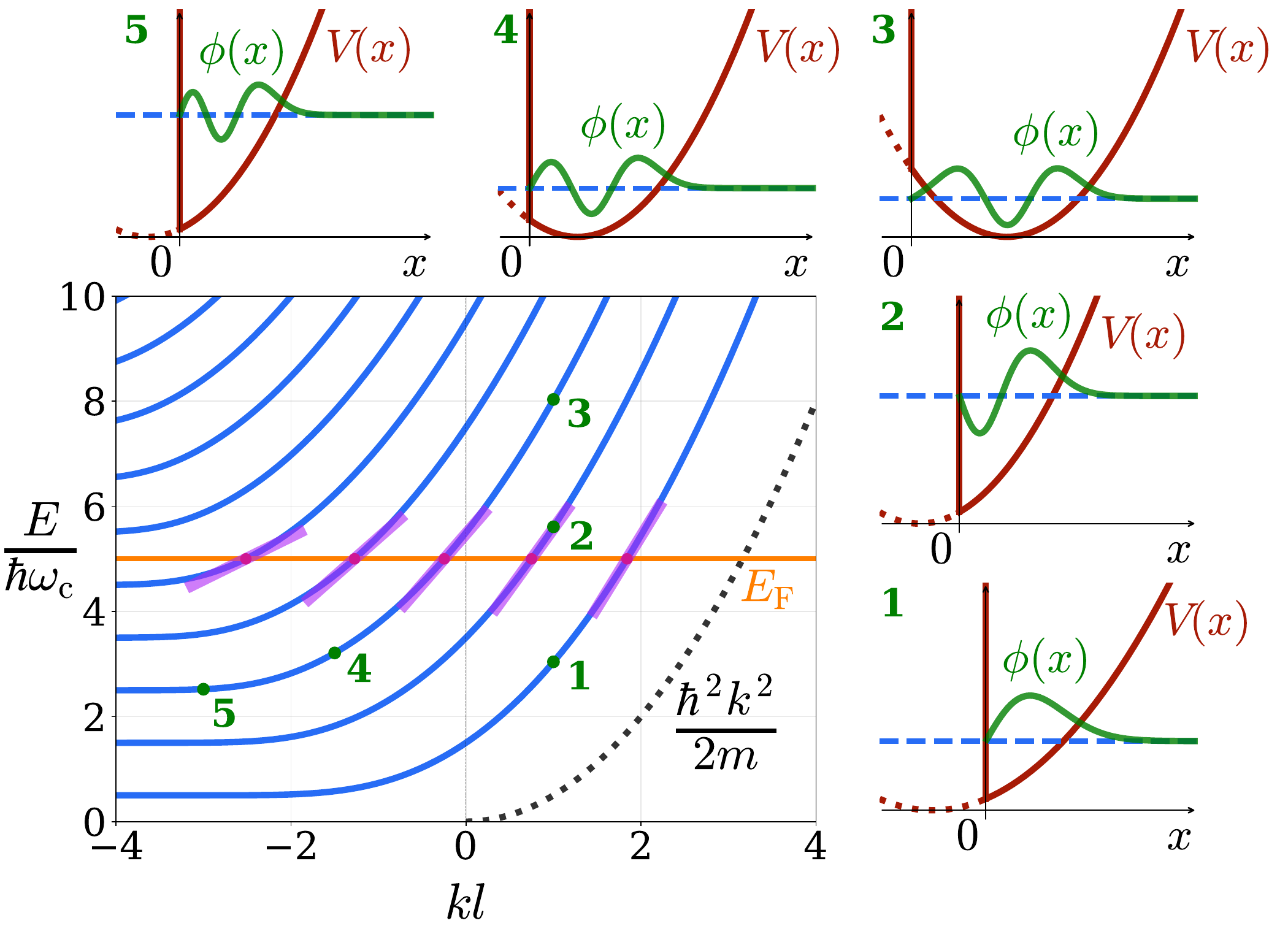}	
	\caption{Main plot: dispersions of the magnetic edge channels at the hard edge of 2DEG (blue curves), and their tangent lines (purple) showing the group velocities when they cross the Fermi level $E_\mathrm{F}$. Top and right insets demonstrate the potential (brown) of the restricted harmonic oscillator and edge-state wave functions (green) at several selected dispersion branches and values of the momentum $k$.}
	\label{Fig2}
\end{figure}

If 2DEG has a hard edge at $x=0$, then the Dirichlet boundary condition $\phi(0)=0$ is applied, which leads to the problem of a harmonic oscillator constrained by an infinitely high potential wall \cite{Dean1966}. The solution of Eq.~(\ref{Schrodinger}) is given by parabolic cylinder functions $U(-E/\hbar\omega_\mathrm{c},\sqrt2(x-x_0)/l)$, thus, the dispersion equation for the energies $E$ of magnetic edge channels as functions of $k$ is
\begin{equation}
U(-E/\hbar\omega_\mathrm{c},\sqrt2lk)=0.\label{2DEG_disp_eq}
\end{equation}
The solutions $E_n$ of Eq.~(\ref{2DEG_disp_eq}) are shown in Fig.~\ref{Fig2}, and their properties are well known \cite{Halperin1982,Dean1966}: at $k\rightarrow-\infty$ the center (\ref{x0}) moves away from the edge, so that the oscillations become asymptotically unrestricted, and the energies tend to those of the bulk Landau levels $E_n^\mathrm{L}=\hbar\omega_\mathrm{c}(n+\frac12)$, $n=0,1,2,\ldots$ When $k$ crosses zero and increases, the restricting wall $x=0$ traverses the oscillation center, giving $E_n$ a dispersion and a positive group velocity $\hbar^{-1}\partial E_n/\partial k$. In the limit $k\rightarrow0$ the wall squeezes the wave function into an almost triangular potential well with the minimum at the energy $m\omega_\mathrm{c}^2x_0^2/2=\hbar^2k^2/2m$, hence the energies follow this quadratic rising trend. If the Fermi level $E_\mathrm{F}$ of 2DEG is located in the gap between the $n$-th and $(n+1)$-th Landau levels, $n+\frac12<E/\hbar\omega_\mathrm{c}<n+\frac32$, then it crosses $n+1$ dispersions of magnetic edge channels. As can be seen in Fig.~\ref{Fig2} the group velocities at the crossing points markedly differ: from the highest for the $n=0$ channel to the lowest for the last branch immediately beneath $E_\mathrm{F}$. All the group velocities do not exceed the Fermi velocity $\sqrt{2E_\mathrm{F}/m}$, which follows from the quadratic trend $E=\hbar^2k^2/2m$ shown by dotted curve.

\begin{figure}[t]
	\centering 
    \includegraphics[width=\columnwidth]{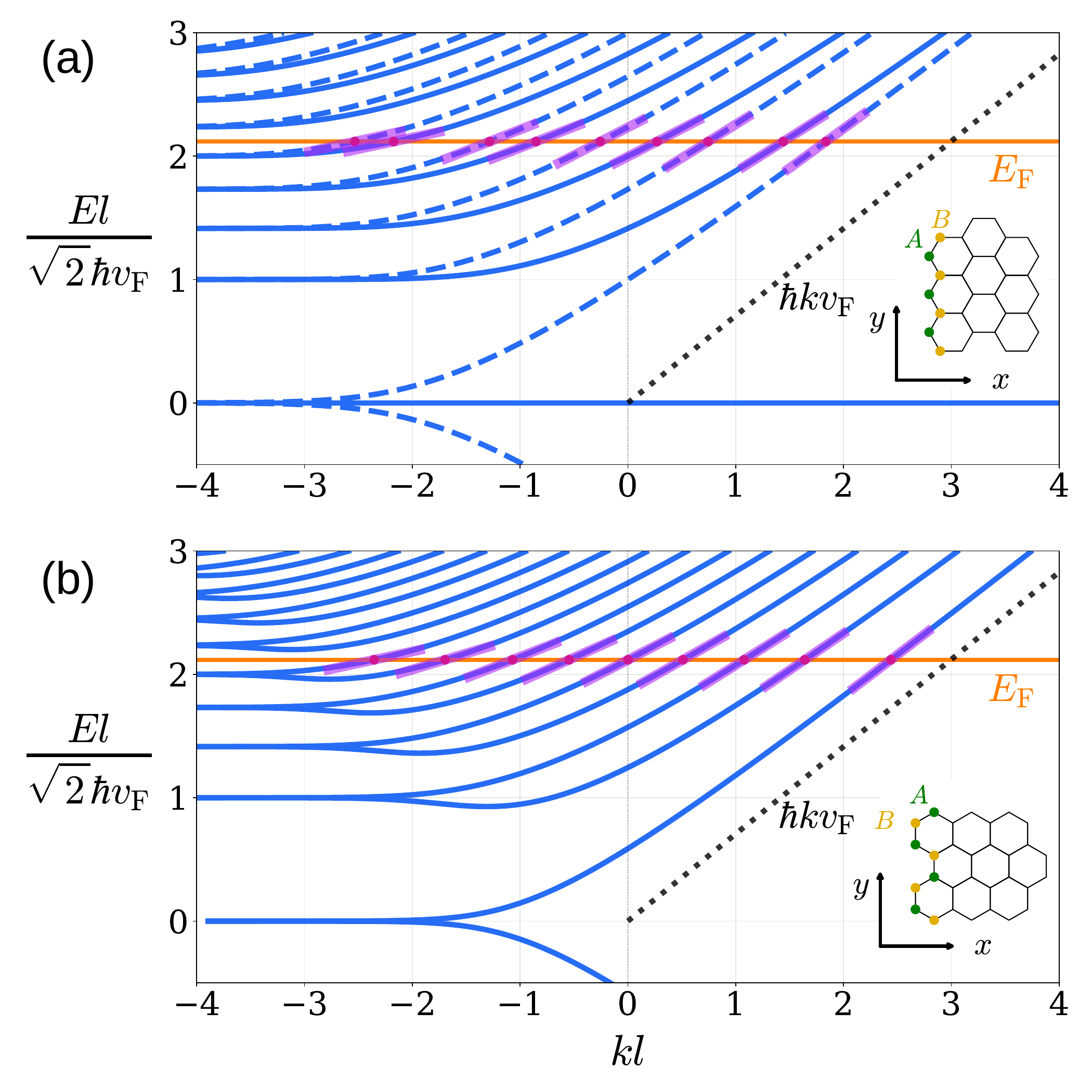}	
	\caption{Dispersions of the magnetic edge channels in graphene for the zigzag (a) and armchair (b) edges (blue curves). Solid and dashed lines in (a) correspond to the valleys $\mathbf{K}$ and $\mathbf{K}'$, while in (b) both valleys are intermixed. The tangent lines (red) show the group velocities at the Fermi level $E_\mathrm{F}$.}
	\label{Fig3}
\end{figure}

\subsection{Graphene}

Massless Dirac electrons in graphene exhibit a behavior which is quite different from those of massive electrons in 2DEG \cite{CastroNeto2009}, however the problem of their magnetic edge channels can likewise be solved analytically \cite{Brey2006,Abanin2007,Gusynin2008}. The energies and wave functions depend on the type of graphene edge (zigzag or armchair) because it determines a boundary condition for a multi-component electron wave function. For the zigzag edge, the dispersion equation reads \cite{Brey2006,Abanin2007,Gusynin2008}
\begin{equation}
U(-\epsilon^2\pm1/2,\sqrt2lk)=0\label{zigzag_disp_eq}
\end{equation}
(besides the additional level $E=0$ in the $\mathbf{K}$ valley), where $\pm$ refers to the valleys $\mathbf{K}$ and $\mathbf{K}'$ respectively, $\epsilon=El/\sqrt2\hbar v_\mathrm{F}$, $v_\mathrm{F}$ is the Fermi velocity. For the armchair edge, electron states in the two valleys are intermixed, and their energies are given by the combined equation
\begin{equation}
\epsilon^2U^2(-\epsilon^2+1/2,\sqrt2lk)-U^2(-\epsilon^2-1/2,\sqrt2lk)=0.\label{armchair_disp_eq}
\end{equation}

The solutions $E_n$ of Eqs.~(\ref{zigzag_disp_eq}) and (\ref{armchair_disp_eq}) are shown in Fig.~\ref{Fig3}. The spectrum of the edge states is particle-hole symmetric and, in the limit $k\rightarrow-\infty$, tends to the bulk Landau levels of graphene \cite{CastroNeto2009} $E_n^\mathrm{L}=\mathrm{sgn}(n)\sqrt{2|n|}\hbar v_\mathrm{F}/l$, $n=0,\pm1,\pm2,\ldots$ In the limit $k\rightarrow\infty$, the dispersions become linear, $E_n\sim\mathrm{sgn}(n)\hbar v_\mathrm{F}k$, therefore their group velocities tend to $v_\mathrm{F}$ and $-v_\mathrm{F}$ in the conduction and valence band, respectively. If the Fermi level is located in the conduction band ($E_\mathrm{F}>0$) between the $n$-th and $(n+1)$-th bulk Landau levels, $E^\mathrm{L}_n<E_\mathrm{F}<E^\mathrm{L}_{n+1}$, it crosses $2n+1$ magnetic edge channels (if the small back-bending of the dispersions is ignored for the armchair edge). Similar to the case of 2DEG, the group velocities of edge channels at the Fermi level vary from the maximal value for the lowest branch to almost zero for the last branch lying immediately below $E_\mathrm{F}$. All the group velocities in Fig.~\ref{Fig3} are bounded from above by the Fermi velocity $v_\mathrm{F}$ which follows from the dispersion of freely propagating massless electrons $E=\hbar v_\mathrm{F}k$ (dotted line).

\subsection{Drift velocity}

Whenever several edge channel dispersions $E_n(k)$ intersect the Fermi level, all of them contribute to the transport and thermodynamic properties of the sample. Consider, for example, the density of states per unit length of the edge:
\begin{equation}
\begin{split}
\rho(E)=\frac1{L_y}\sum_{nk}(E-E_n(k))&=\frac1{2\pi}\sum_n\int dk\:\delta(E-E_n(k))\\
&=\frac1{2\pi}\sum_n\frac1{\partial E_n/\partial k|_{E_n=E}}.
\end{split}
\end{equation}
Here we have switched from summation over $k$ to integration in the limit $L_y\rightarrow\infty$, and the last sum contains only those channels $n$ that intersect the Fermi level. Noting that
\begin{equation}
v_n=\frac1\hbar\frac{\partial E_n}{\partial k}\label{v_n}
\end{equation}
is the group velocity of the $n$-th edge channel along the edge, one arrives at the summation of their inverses:
\begin{equation}
\rho(E)=\frac1{2\pi\hbar}\sum_n\frac1{v_n}.
\end{equation}
This formula provides the contribution to the density of states from the edge channels in the gaps between bulk Landau levels \cite{Buttiker1985}.

Defining the drift velocity of the magnetic edge channels as harmonic mean of the group velocities at the Fermi level
\begin{equation}
v_\mathrm{D}=\left(\frac1N\sum_n\frac1{v_n}\right)^{-1}\equiv\langle v_n^{-1}\rangle^{-1},\label{v_D_def}
\end{equation}
where $N$ is the number of channels in the sum, gives 
\begin{equation}
\rho(E)=\frac{N}{2\pi\hbar v_\mathrm{D}}.
\end{equation}
Furthermore, other thermodynamic and electric properties of the sample, such as compressibility $\varkappa=-V^{-1}(\partial V/\partial p)=n_\mathrm{e}^{-2}(\partial n_\mathrm{e}/\partial\mu)$ and quantum capacitance $C_\mathrm{Q}=e^2(\partial n_\mathrm{e}/\partial\mu)$ of an electron gas, where $n_\mathrm{e}$ is the electron density, and $\mu$ is the chemical potential, can be related to the density of states and thus to the drift velocity, because $\partial n_\mathrm{e}/\partial\mu=g\rho(E_\mathrm{F})$ at low temperatures and with neglecting electron interaction effects ($g$ is the electron state degeneracy factor).

On the other hand, the total current carried by the edge channels is (taking into account the negative electron charge) 
\cite{Halperin1982,Buttiker1985}
\begin{equation}
I=-\frac{eg}{2\pi\hbar}\sum_n\int_{-\infty}^{k_n}dk\:\frac{\partial E_n}{\partial k}=-\frac{eg}{2\pi\hbar}\sum_n(E_\mathrm{F}-E_n^\mathrm{L}),
\end{equation}
where $k_n$ is the solution of $E_n(k_n)=E_\mathrm{F}$, and only those edge channels $n$ that are partially occupied (with $E_n^\mathrm{L}<E_\mathrm{F}$) enter the sum. Variations of the local Fermi level at the edge induce the current variations $\partial I/\partial E_\mathrm{F}=-egN/2\pi\hbar$. Therefore, the response of $I$ on variations of the charge density per unit length accumulated at the edge $Q=-en_\mathrm{e}$ is exactly the drift velocity \cite{Sokolik2024}:
\begin{equation}
\frac{\partial I}{\partial Q}=\frac{\partial I/\partial E_\mathrm{F}}{\partial Q/\partial E_\mathrm{F}}=v_\mathrm{D}.\label{dIdQ}
\end{equation}
Thus, the sample edge in a magnetic field behaves as an element of a distributed circuit with quantum capacitance per unit length $C_\mathrm{Q}=e^2N/2\pi\hbar v_\mathrm{D}$ and a current response along the edge to accumulated charge (\ref{dIdQ}), both related to the harmonically averaged drift velocity (\ref{v_D_def}). The response (\ref{dIdQ}) enters the equations describing edge magnetoplasmons, raising their velocity by an amount of $v_\mathrm{D}$ \cite{Sokolik2024} or by some typical value of edge channel group velocity \cite{Balev1997}. The last result can also be obtained by considering the edge electrostatics in terms of distributed circuits \cite{Hashisaka2013,Lin2021} or starting from a quantum description of edge channels and their self-consistent electrostatic response \cite{Volkov1991,Mikhailov2001,Bosco2017,Wang2023}.

\section{Semiclassical analysis}
\label{Sec3}

\subsection{Isotropic dispersion}

Now consider the semiclassical limit where the number of edge channels $N$ crossing the Fermi level is large, and thus we can refer to a classical motion of an electron along the skipping orbits. With isotropic and monotonous electron dispersion, each cyclotron orbit is circular, with the radius $r$ and angular frequency $\omega_\mathrm{c}$ unambiguously determined by the energy $E$. In particular, for massive electrons with the dispersion $E=p^2/2m$ it is $\omega_\mathrm{c}=\hbar/ml^2$, $r=l^2\sqrt{2mE}/\hbar$, and for massless electrons in graphene with the linear dispersion $E=v_\mathrm{F}p$ we get+ $\omega_\mathrm{c}=\hbar v_\mathrm{F}^2/El^2$, $r=El^2/\hbar v_\mathrm{F}$.

With the half span angle $\theta$ of the skipping orbit (Fig.~\ref{Fig1}), the $x$ coordinate of its center is $x_0=-r\cos\theta$. The distance $2r\sin\theta$ between two successive collisions with the edge is traversed at time $2\theta/\omega_\mathrm{c}$, thus the average $y$-velocity of an electron is
\begin{equation}
v=v_0\frac{\sin\theta}\theta,\label{v_classical}
\end{equation}
where $v_0=r\omega_\mathrm{c}$ is the electron linear velocity at the cyclotron orbit.

The quantization of skipping orbits requires an integer number of magnetic flux quanta passing through each arc, thus, its area $S=r^2(\theta-\sin\theta\cos\theta)$ should be equal to $2\pi l^2(n+\gamma)$, where $n=0,1,2,\ldots$, and $\gamma$ is the Morse-Maslov index related to caustics and other peculiar points of the trajectory \cite{Brack1997}. Alternatively, one can semiclassically quantize \cite{Avishai2008,Montambaux2011} one-dimensional particle motion along the $x$ axis in the quadratic potential of Eq.~(\ref{Schrodinger}), namely $V(x)=\hbar\omega_\mathrm{c}(x-x_0)^2/2l^2$, constrained by a hard wall at $x=0$: $\int_0^{x_\mathrm{r}}p(x)\:dx=\pi\hbar(n+\gamma)$. Here $x_\mathrm{r}$ is the right turning point where $V(x_\mathrm{r})=E$, and the function $p(x)$ depends on the electron dispersion. Both approaches result in the same quantization condition
\begin{equation}
\theta-\sin\theta\cos\theta=\frac{2\pi l^2(n+\gamma)}{r^2},\label{quant1}
\end{equation}
if the connection between $E$, $\omega_\mathrm{c}$, and $r$ is taken into account.

Therefore, the harmonic averaging (\ref{v_D_def}) of the velocities (\ref{v_classical}) should be performed over the quantized numbers $n$ of the edge channels at a fixed energy $E$ equal to the Fermi energy $E_\mathrm{F}$. If the number of channels $N$ is large, one can switch from summation to integration and obtain
\begin{equation}
v_\mathrm{D}^{-1}\approx\frac{\int_0^Ndn\:v^{-1}}{\int_0^Ndn}.\label{v_D_int}
\end{equation}
Since $E=\mathrm{const}$ during the integration, $r$ is constant as well, and  Eq.~(\ref{quant1}) yields $dn=(r^2\sin^2\theta/\pi l^2)d\theta$. Using Eq.~(\ref{v_classical}), one derives
\begin{equation}
v_\mathrm{D}^{-1}=v_0^{-1}\frac{\int_0^\pi d\theta\:\theta\sin\theta}{\int_0^\pi d\theta\:\sin^2\theta}=\frac2{v_0}.\label{v_D_inv}
\end{equation}
Since the velocity $v_0$ for all edge channels at the Fermi level equals the Fermi velocity $v_\mathrm{F}$, we obtain our main result
\begin{equation}
v_\mathrm{D}=\frac12v_\mathrm{F}.\label{v_D_result}
\end{equation}
It can be compared with the arithmetic mean of edge channel group velocities, which was proposed as an estimate for the edge magnetoplasmon drift velocity by some authors \cite{Volkov1991,Mikhailov2001,Petkovic2014}:
\begin{equation}
\langle v_n\rangle=\frac1N\sum_n v_n\approx\frac{\int_0^Ndn\:v}{\int_0^Ndn}
=v_\mathrm{F}\frac{\int_0^\pi d\theta\:\sin^3\theta/\theta}{\int_0^\pi d\theta\:\sin^2\theta}\approx0.62v_\mathrm{F}.
\end{equation}
As expected, it is larger than the harmonic mean, although not significantly.

\subsection{Anisotropic dispersion}

The result (\ref{v_D_result}) can be obtained using a more general semiclassical quantization that is not limited to circular orbits. Consider an electron with anisotropic dispersion moving along a skipping orbit of arbitrary shape, Fig.~\ref{Fig1}(c). The only requirement we impose is the $y\rightarrow-y$ symmetry of the full (uninterrupted) cyclotron orbit, which allows preservation of its shape after reflection from the edge. In this case, the skipping orbit consists of mutually $y$-shifted parts of the full cyclotron orbit restricted to the region $x>0$. In the absence of such symmetry, the skipping orbit can undergo chaotization due to incompatibility of the orbit and boundary symmetries \cite{Schoeppl2025}, making our analysis inapplicable. We define $x_0$ as the distance between some reference point of the cyclotron orbit and the edge. The action between two successive collisions with the edge
\begin{equation}
S(E,x_0)=\oint p_x(x)dx
\end{equation}
depends on $E$ and $x_0$, which determine the location of the left turning point. The semiclassical quantization condition
\begin{equation}
S(E,x_0)=2\pi\hbar(n+\gamma)\label{quant2}
\end{equation}
imposes a relationship between $E$, $n$, and $x_0$.

From the gauge invariant minimal substitution $\mathbf{p}\rightarrow\mathbf{p}+e\mathbf{A}/c$, the universal relation (\ref{x0}) between $x_0$ and momentum $k$ along the edge follows (up to an additive constant) that the group velocity of a specific edge channel is $v_n=\hbar^{-1}(\partial E/\partial k)_n=-l^2\hbar^{-1}(\partial E/\partial x_0)_n$. Since $E$ and $x_0$ should be changed simultaneously to maintain the constant action (\ref{quant2}) at $n=\mathrm{const}$, for the inverse group velocity, one finds:
\begin{equation}
\frac1{v_n}=\frac\hbar{l^2}\frac{(\partial S/\partial E)_{x_0}}{(\partial S/\partial x_0)_E}.
\end{equation}
In contrast, the averaging (\ref{v_D_int}) over edge channel numbers $n$ is carried out at $E=E_\mathrm{F}=\mathrm{const}$, when $dn=(dx_0/2\pi\hbar)(\partial S/\partial x_0)_E$, which gives the expression $v_\mathrm{D}^{-1}=(2\pi l^2 N)^{-1}\int dx_0\:(\partial S/\partial E)_{x_0}$, where $(\partial S/\partial x_0)_E$ is canceled. As known, the derivative of action by energy is the period of oscillations \cite{Landau1976}, thus
\begin{equation}
v_\mathrm{D}^{-1}=\frac1{2\pi l^2N}\int dx_0\:T(x_0),\label{v_D_gen1}
\end{equation}
where $T(x_0)$ is the duration of cyclotron motion at the Fermi energy between two successive collisions with the edge, dependent on the $x$-coordinate of the skipping orbit $x_0$. Measuring the coordinate $x'=x+d_1-x_0$ from the left end of the cyclotron orbit (Fig.~\ref{Fig1}(c)), one can write $T(x_0)=2\int_{d_1-x_0}^{d_1+d_2}dx'/v_x(x')$, where $v_x(x')$ is the $x$-projection of the electron velocity along the cyclotron orbit at the coordinate $x'$. Thus, it leads to the double integral where the order of integration can be changed:
\begin{equation}
\begin{split}
\int dx_0\:T(x_0)&=2\int_{-d_2}^{d_1}dx_0\int_{d_1-x_0}^{d_1+d_2}\frac{dx'}{v_x(x')}\\
&=2\int_0^{d_1+d_2}\frac{x'dx'}{v_x(x')}=\int_0^{T_0}x'dt=T_0\langle x'\rangle_t,
\end{split}
\end{equation}
where $T_0$ is the period of the full cyclotron orbit at the Fermi level, and $\langle\ldots\rangle_t$ is the time average along it.

On the other hand, the semiclassical quantization of the cyclotron orbit implies $2\pi l^2N\approx S_0$ (neglecting the Morse-Maslov index at $N\gg1$), where the area inside the cyclotron orbit can be written as an integral along it: $S_0=\oint x'dy=\int_0^{T_0}x'v_ydt=T_0\langle x'v_y\rangle_t$. Thus Eq.~(\ref{v_D_gen1}) transforms into our second main result:
\begin{equation}
v_\mathrm{D}=\frac{\langle x'v_y\rangle_t}{\langle x'\rangle_t}.
\end{equation}
This formula means that the drift velocity is given by a degree of correlation between the $x'$ coordinate and the velocity $v_y$ along the cyclotron orbit, where $x'$ is measured from its point closest to the edge. For a circular orbit of radius $r$, we obtain $\langle x'v_y\rangle_t=\frac12rv_0$, $\langle x'\rangle_t=r$, and the result (\ref{v_D_inv}) is reproduced. For anisotropic orbits, $v_\mathrm{D}$ can be larger (smaller) than the half of an average cyclotron velocity if $|v_y|$ is higher (lower) on average in its far-from-the edge part than in close-to-the edge one.

\begin{figure}[t]
	\centering 
	\includegraphics[width=\columnwidth]{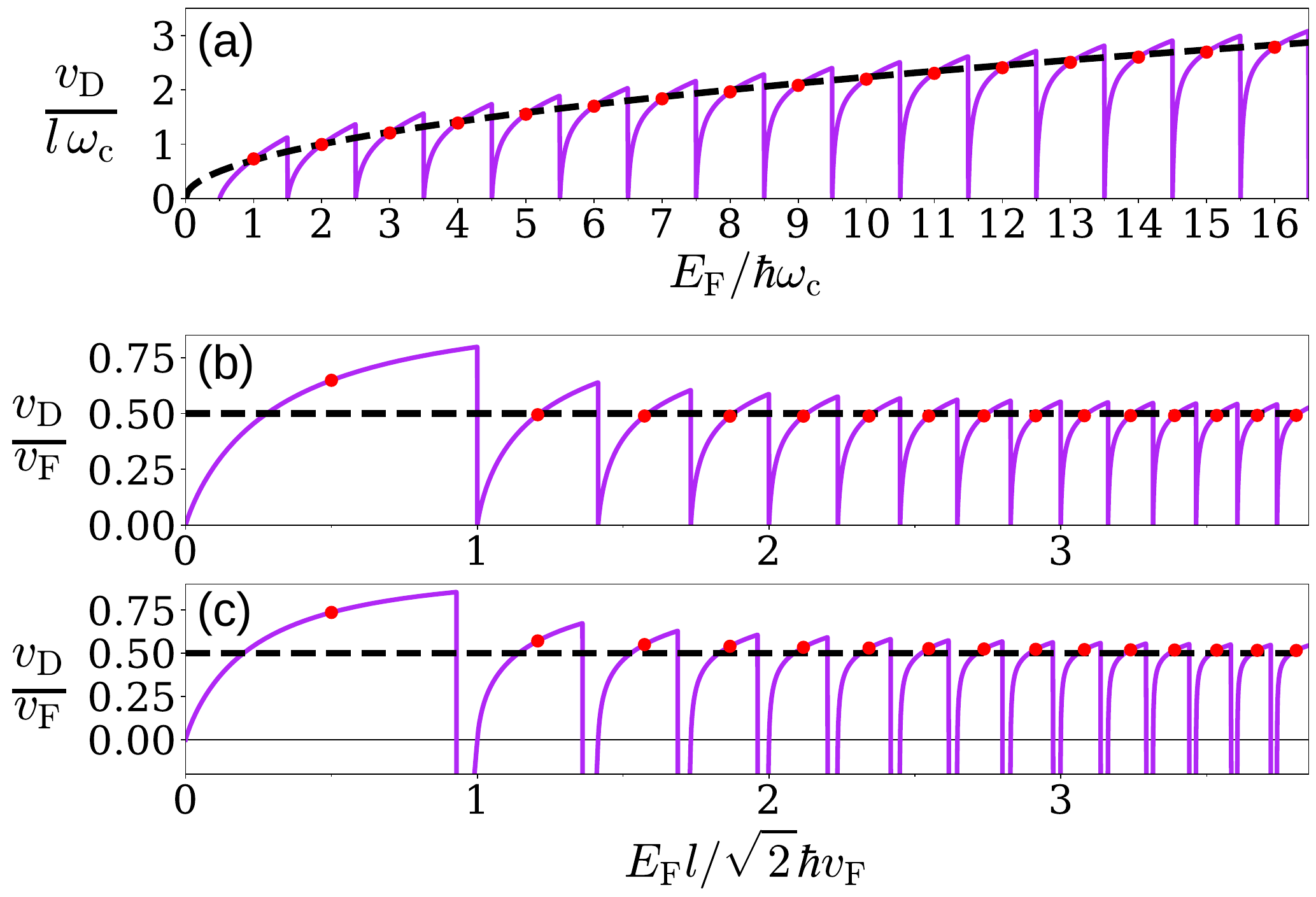}	
	\caption{Drift velocities for 2DEG (a), and graphene with (b) zigzag and (c) armchair edges as functions of the Fermi energy. Dashed lines show the semiclassical limit $v_\mathrm{D}=\frac12v_\mathrm{F}$, and red points mark the Fermi energies in the mid-points between bulk Landau levels $E_n^\mathrm{L}$.}
	\label{Fig4}
\end{figure}

\section{Discussion}
\label{Sec4}

We checked numerically the semiclassical estimates (\ref{v_D_result}) of the drift velocity for 2DEG and graphene. With calculating the group velocities from the dispersion equations (\ref{2DEG_disp_eq}), (\ref{zigzag_disp_eq}), or (\ref{armchair_disp_eq}), we took their harmonic mean (\ref{v_D_def}) and plotted the resulting $v_\mathrm{D}$ as functions of $E_\mathrm{F}$ in Fig.~\ref{Fig4}. The drift velocities demonstrate quantum oscillations, reflecting the discreteness of Landau levels $E_n^\mathrm{L}$. In Figs.~\ref{Fig4}(a,b), $v_\mathrm{D}$ drops to zero at $E_\mathrm{F}\rightarrow E_n^\mathrm{L}+0$, because the group velocity of the highest occupied $n$-th edge channel transforming into the bulk Landau level tends to zero and dominates the harmonic mean. In the case of graphene with armchair edge, Fig.~\ref{Fig4}(c), $v_\mathrm{D}$ becomes negative in narrow ranges below each $E_n^\mathrm{L}$ due to dispersion back-bendings (see Fig.~\ref{Fig3}(b)), providing dominating negative contributions of the upstream branch to the harmonic mean. As characteristic values of $v_\mathrm{D}$ which most closely correspond to the classical regime, we assume those at the mid-gaps between Landau levels $E_\mathrm{F}=\frac12(E_n^\mathrm{L}+E_{n+1}^\mathrm{L})$ when the system resides in the quantum Hall insulating regime, and intra-Landau level excitations are maximally suppressed. As seen in Fig.~\ref{Fig4}, the drift velocity at these points tends to the semiclassical limit $\frac12v_\mathrm{F}$ at $n\rightarrow\infty$, which agrees with our analytical estimate.

In this paper, the hard edge of a 2D electron system was considered, which naturally appears in the case of 2D crystals and can be made by a chemical etching of semiconducting quantum wells \cite{Huber2005}. However a lot of work was devoted to a study of smooth edges of 2DEG created by electrostatic gates, complex electrostatics of magnetic edge channels in smooth confining potentials, and their possible reconstructions \cite{Chklovskii1992,Coissard2023,Yu2025}. Let us assume a smooth potential $U(x)$ confining electrons to the region $x>0$ instead of a hard edge, neglecting electron interactions and the screening of this potential by the 2DEG itself. In this case, if the length scale of nonlinearity of $U(x)$ is small compared to $l$, the edge channel energies would be $E_n=E_n^\mathrm{L}+U(-l^2k)$, i.e. bulk Landau level energies are shifted by the confining potential, which grows in the negative $x$ direction. The group velocities (\ref{v_n}) are determined by the steepness of the confining potential, $v_n=-l^2\hbar^{-1}U'(-l^2k)$. If the latter is created by a uniform electric field $\mathcal{E}$, i.e. $U=-e\mathcal{E}x\,\Theta(-x)$, then all $v_n$ and $v_\mathrm{D}$ are equal to $\mathcal{E}c/B$. It is the well-known drift velocity of electrons undergoing a spiraling motion in crossed electric and magnetic fields, and some authors interpret the edge magnetoplasmon drift velocity from such classical perspective \cite{Wassermeier1990,Talyanskii1992,Hashisaka2013,Petkovic2014}. However, when $U(x)$ is a nonlinear function, $v_\mathrm{D}$ depends on its specific form and on a specific electron dispersion. In both linear and nonlinear cases, $v_\mathrm{D}$ for a smooth edge is not related to $v_\mathrm{F}$, so the universal result (\ref{v_D_result}) does not apply.

To conclude, we have emphasized the crucial role of the harmonic mean of magnetic edge channels group velocities at the Fermi level, which affects the density of states, quantum capacitance, and the electron gas current response near the edge in a magnetic field. This mean value, termed the drift velocity, contributes to the total velocity of edge magnetoplasmons and was estimated in several experiments as being of the order of the Fermi velocity. Using semiclassical quantization, we have shown that the drift velocity is exactly 1/2 of the Fermi velocity for hard edge and any isotropic dispersion, when the large number of bulk Landau levels is occupied. This result have been confirmed by numerical calculations for 2DEG and graphene with two types of the edge. For anisotropic dispersions, the closed-form expression for the drift velocity in terms of the correlation between electron coordinate and velocity in perpendicular directions along the cyclotron orbit was obtained. These results can be used for more accurate and informed interpretations of experiments on edge magnetoplasmons and on electron properties of magnetic edge channels.

Several directions of the further studies can be drawn. First, we have considered only circular or mirror-symmetric cyclotron orbits colliding with a straight edge, when the skipping orbit has a self-repeating and predictable shape. It is of interest to study the scenario in which these conditions are violated, and the skipping orbits become chaotic \cite{Schoeppl2025}. Second, we have assumed a monotonous electron dispersion, when a single cyclotron orbit corresponds to each energy, and all skipping orbits propagate in the same downstream direction. An appearance of upstream modes due to a nonmonotomous electron dispersion \cite{Erhardt2025} or interaction effects \cite{Bid2010} can significantly alter the picture. Third, our semiclassical analysis is rather crude and neglects the role of Morse-Maslov index, Berry phases, and discreteness of edge channels, thus, more accurate semiclassical theory of edge states in 2DEG \cite{Avishai2008,Montambaux2011} and graphene \cite{Delplace2010} based on WKB approximation might provide refined results. Fourth, how the drift velocity changes in the presence of disorder, which disrupts the skipping orbits through electron scattering at the semiclassical level or creates charge puddles in the electrostatic picture \cite{Lin2021}, is also worth an investigation. Finally, future work should address the comparison between the considered hard-edge drift velocity analysis and the case of smooth electrostatically created edges of 2DEG, where multiple branches of edge magnetoplasmons with different transverse field profiles \cite{Mikhailov2001,Aleiner1994} appear, and edge channels are defined by a sequence of compressible and incompressible stripes \cite{Chklovskii1992}.

\section*{Acknowledgements}
The work was supported by the Russian Science Foundation (Grant No. 23-12-00115-P). E.R.S. acknowledges support for her work on numerical calculations from the Program of Basic Research of the Higher School of Economics.

\bibliographystyle{elsarticle-num} 
\bibliography{references}

\end{document}